# Dis-embedded Openness:
# Inequalities in European Economic Integration at the Sectoral Level

Balazs Vedres
Central European University

Carl Nordlund
Central European University; Lund University

The process of European integration resulted in a marked increase in transnational economic flows, yet regional inequalities along many developmental indicators remain.  We analyze the unevenness of European economies with respect to the embedding of export sectors in upstream domestic flows, and their dependency on dominant export partners.  We use the WIOD data set of sectoral flows for the period of 1995-2011 for 24 European countries.  We found that East European economies were significantly more likely to experience increasing unevenness and dependency with increasing openness, while core countries of Europe managed to decrease their unevenness while increasing their openness.  Nevertheless, by analyzing the trajectories of changes for each country, we see that East European countries are also experiencing a turning point, either switching to a path similar to the core, or to a retrograde path with decreasing openness.  We analyze our data using pooled time series models and case studies of country trajectories.

**Introduction**

The economic integration of EU member states is a central element of the European project, where the standardization of regulations, a customs union, the removal of institutional barriers were designed to facilitate the emergence of a larger coherent European economic unit (Balassa 1962). According to these expectations, a high-degree of economic integration eventually erases the preferentiality of economic exchange (along lines of nationality, language, tradition).  Flows will reflect, it is assumed, only the rationalities of space, quality, and cost.

The European process of economic integration did result in an increase of flows, according to analyses of bilateral trade data and other aggregate indicators at the national and regional levels (Hoen 2002; Bergstrand 2008)[1].  This is particularly evident for the new Eastern countries: with the removal of tariff and non-tariff barriers liberating the movement of goods, capital and services, coupled with the prescribed institutional and regulatory harmonization, flows increased greatly between the economies of the older European core and the new East European member states. East-European economies were not only connected to Core markets, but sectors from the East became integrated into European-wide production structures as well.

---

[1] Although there is an overall agreement of the trade-creating effects of economic integration, particularly with respect to the European case (e.g. Bergstrand 2008), conceptualizing and measuring such an effect is not a trivial exercise. Whether comparing intra- and extra-area trade or extrapolated pre-integration data with actual post-integration observations, such ex-post assessments, similar to pre-integration assessments of would-be effects, are inherently  (Balassa 1967).

However, the European project is not only about the economic benefits of increasing flows – better economies of scale, employment, higher profits. From the perspective of broader developmental concerns, increased economic transnationalism might preserve or even amplify deeper inequalities. While European economies are becoming increasingly integrated with increased transnational flows, gaps in welfare, wages, factor costs and value-added productivity that separate the East from the West seem to persist.

One interpretation takes the lack of convergence to be a transient phenomenon, akin to a Kuznets-curve of economic integration where "close integration is good, but a limited move towards integration might hurt" (Krugman 1991:89). A second interpretation for gaps in developmental indicators is that these stem from durable core-periphery relations. According to this view, economic integration is not a source of increasing equality, but rather the cause of structural and sectorial imbalances. As integration increases, economies on the periphery are locked into vertical trade, foreign-dominated consumer markets, and low value-added positions in the global chains of production (e.g. Oman and Wignaraja 1991; So 1990). According to the more recent version of this argument, adjusted to the growing centrality of global value chains (GVC) in the new Eastern member states, a new type of dependent market economy (DME) has emerged in these countries with the headquarters of the multinational firms capitalizing on the Eastern cheap and highly skilled labor and, keeping the positions of firms from these countries at the low value added ends of the production chains. Finally, according to a third interpretation, exposed also in the introduction to the special issue, the new member states dramatically differ from each other in the form and strengths of domestic developmental agency, in the capacity of domestic public and private actors to shape developmental paths (Bohle and Greskovits 2012; Bruszt et al, 2015). Based on this third approach one would expect diverging developmental pathways among the Eastern member states. Empirical research thus far did not produce decisive evidence on the developmental effects of the spread of region-wide production chains in Europe for any of the above the arguments.

We fill this gap in this paper. Instead of measuring underdevelopment as just a singular dimension of the "not there yet", it is more fruitful to analyze pathways – divergent and path dependent processes of institution building and economic development – that European peripheries took (Bohle and Greskovits 2012). Beyond aggregate statistics and overall correlations we also analyze trajectories of European economies to find evidence for convergence, divergence, or durable regional inequality.

A key form of inequality in economic integration stems from the way in which transnationally integrated economic activity is embedded domestically. A risk seen in increased transnational flows is the production of disembedding (Scott 1997), where transnationalization takes the form of cathedrals in the desert (Hardy 1998): places of transnational production increasingly disconnected from domestic structures. Research about the diverse ways in which the automotive industry became integrated on the European peripheries indicate that the depth of domestic embedding of manufacturing sectors, such as transport equipment manufacturing, is a key factor in the success of economic integration (Bruszt et al 2015). Even if institutional structures are congruent, incongruent supply structures in Eastern Europe might block the success of transnational integration (Greskovits 2005).

Such disembedding can lead to sustained under-development by preventing material benefits from transnational participation from reaching a wider part of the economy.  If an economy relies on export sectors that are dis-embedded from the domestic sectoral flows, the benefits of increased exports will be limited to the export sectors themselves, leading to stagnation in other sectors and resulting in problems of economic dualism (e.g. Singer 1970).  Disembedding can be detrimental by blocking ties of learning, both of know-how related to production processes, and both of knowledge about market opportunities (Maya-Ambia 2011).  It might leave an economy vulnerable to the flight of capital (as production facilities in dis-embedded sectors might be more easy to relocate), and it might also lead to the strong bargaining position of industries in transnationally embedded sectors (by, for example, credible threats of relocation).

In this article we analyze the process of European economic integration along two main dimensions: economic openness and the domestic embedding of transnationalized production. We also analyze the degree of trade partner concentration of export-oriented sectors. Our empirical approach is based on input-output tables of economic sectors.  Our basic unit of analysis is a European economic sector. Based on the World Input-Output Data project we use data on flows among 816 sectors (34 sectors in each of 24 national economies) over the period of 1995-2011. We develop three metrics: transnational openness, unevenness in the domestic upstream embedding of export sectors, and dependency on dominant export partners.  We relate the openness of European economies to their unevenness in terms of sectoral embedding to identify how increasing openness is related to unevenness.  Our analysis operates at three levels: at the level of particular sectors, at the level of national economies, and at the level of larger regions.

Our first measure – transnational openness – captures, for each country, the ratio between inter-sectorial value flows that cross the national borders and those that are domestic. Not surprisingly, the results from our dataset reflect previous assessments: although the starting points differ, the national industrial sectors in Europe are becoming increasingly more connected with sectors in other countries[2].

Our second measure is upstream domestic embedding (defined at the level of sectors), and the uneven distribution of this embedding (defined at the level of a national economy).  This measure captures the difference between a sector's share of total exports and its share as user of intermediate domestic inputs. This allows us to identify sectors that are significant exporters, but are weakly embedded in domestic upstream flows.  Coupling this with the share of sectorial import allow us to identify sectors that are "cathedrals in the desert" and, through the distributional variance of these values, to see the overall unevenness of a national economy.

Looking beyond domestic production structures, our third metric captures sectorial export-dependency in terms of foreign partner concentration of sectors. Primarily associated with dependency theory and related studies on the developmental effects of partner concentration (e.g. Galtung 1971; Dominguez 1971; Berman 1974; Chan 1982), the topology of international patterns of exchange and would-be monopolistic-oligopsonic patterns of exchange are equally relevant to understand market access (e.g. Condliffe 1950:816; Bauer 1954:103; Meier and Baldwin 1957:332), configurations of global commodity/value chains (Wallerstein and Hopkins 2000 [1985]; Porter 1987; Gereffi and Korzeniewicz 1994) and the interplay between such topologies and development (e.g. Heintz 2006:515; Appelbaum et al 1994). To cater for the differences in relative sizes between countries, our measure captures the percentage-point difference between the largest and second-largest shares of outflows for each sector in each country and year. We apply this metric on those industrial sectors identified as having a significant transnational openness (i.e. our first metric).

---

[2] Whereas "economic integration" typically refers to the institutional and regulatory processes towards (and state of) the creation of a common market (e.g. Balassa 1962), our usage of the term refers explicitly to cross-border economic exchange.

Our findings indicate that between 1995 and 2011 openness increased almost monotonically in all three regions: The Core, the GIPS (Greece, Ireland, Portugal, and Spain), and the East. There is however a difference between the three regions in terms of unevenness: increasing openness was paired with decreasing unevenness in the Core, while more openness meant more unevenness in the East. The GIPS region was highly diverse in this respect. Dependency shows a similar pattern: increasing openness in the East was related to an increase in dependency, while in the West and GIPS countries dependency is not a function of openness.

The relationship between openness and dependency shows a very similar pattern to the relationship between openness and unevenness. For the Core countries there is no evidence for increasing dependency as their economies are becoming more open, while countries in the East show a significant trend: more openness here means an increase in sectoral dependency on foreign export target sectors. The GIPS region shows a similar trend to the East region, but this trend is not statistically significant.

Beyond estimating the correlation among variables of openness, unevenness, and dependency, we also analyzed the amount of change (trajectories of temporal volatility) that each country experienced. Such trajectories in the East experienced larger changes, than in the Core. At the level of national economies, we found that it was only the Core, where economies were able to increase openness and decrease unevenness at the same time. Trajectories for economies in the GIPS and East regions were much more volatile. We found two distinct kinds of trajectories in the East: turning point and retrograde trajectories. Turning point trajectories were able to reverse the trend of jointly increasing openness and unevenness, and switch onto a path where unevenness decreases with further increase of openness. Retrograde trajectories was decrease in unevenness only when openness also decreased. Trajectories in the GIPS region show the most volatility of all, but in very diverse directions. Greece experienced a dramatic increase in unevenness with hardly any change in openness, while Ireland shows the opposite pattern – a drastic increase in openness with a modest increase in unevenness.

We analyze two cases from each of the three regions – Germany and France from the Core, Greece and Ireland from the GIPS region, and Hungary and Estonia from the East. The case studies highlight the usefulness of using upstream domestic embedding as a dimension to identify sectors that are most related to developmental outcomes in a national economy.

**World Input-Output Data**

An Input-Output table records directional valued flows between (and within) industrial sectors or product groups. Derived from national Supply-Use tables that capture the supply of domestically produced and imported goods and services and their intermediate use, domestic final consumption and exports, a national Input-Output table – see Table 1 - is usually a balanced account, where the data on

intra- and inter-sectorial flows (Z) is supplemented with imports for intermediate (I) and final use (IFU), exports (E), domestic final use (DFU) and various value-added categories – see Table 1 below.

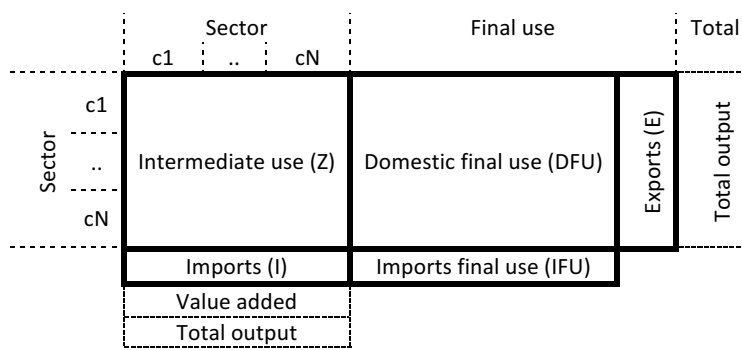

Table 1: General layout of a national Input-Output table (from Timmer 2012, p. 63)

Apart from the Input-Output tables produced by national statistical agencies, there are several data providers that compile and disseminate standardized national Input-output data – such as OECD, Eurostat and World Bank. The data used in this chapter is taken from a recent project, the World Input-Output Data initiative (WIOD for short) (Timmer 2012; Dietzenbacher et al 2013; Timmer et al 2015). WIOD is a multi-regional dataset that merges national Input-Output data with bilateral trade flow statistics. It contains annual Input-Output data for 34[3] sectors between 1995-2011 for 40 countries (including a virtual Rest-of-World country), out of which 27 are EU member states as of 2012.

Analyses of transnationalization typically relies on international trade data that is recorded at the level of national economies. The use of multi-regional Input-Output data such as WIOD allows for a decomposition of national economies into their constituent sectors. This results in a more complex network: for example, trade flows for Europe can be depicted as a network of 24 constituting economies (Croatia, Cyprus, Luxemburg, and Malta are excluded from our analysis here of the 28 EU countries for their missing data, small size, and uniqueness). European economic flows can also be depicted as flows among 816 sectors (34 for each of the 24 countries). This network opens the possibility to address inequalities between economies stemming from the structure of flows within and outside countries at the sectoral level.

Our article uses this data to connect domestic sectoral flows (or the absence of domestic flows) to outside flows to sectors in other countries, to compare openness of the economy and the domestic upstream embeddedness of export sectors. As our questions are concerned with the state of the pan-European production structures, our analysis focuses on the intermediate use sections of the national Input-Output tables (Z in Table 1) as well as the vector of cross-border exports (E) and imports (I). Covering annual domestic intra- and inter-sectorial flows, final domestic use as well as exports and imports for 34 sectors between 1995-2011, the data covers 24 EU countries that we have separated into three subsets: Core (Austria, Belgium, Germany, Denmark, Finland, France, Great Britain, Italy, the Netherlands, Sweden), GIPS (Greece, Ireland, Portugal, Spain), and East (Bulgaria, Czech Republic, Estonia, Hungary, Lithuania, Latvia, Poland, Romania, Slovakia, Slovenia).

---

[3] WIOD uses a sectorial nomenclature comprising 35 sectors, but as there is no data on intermediate flows for the 'Private Households with Employed Persons' (c35) sector, this sector is excluded in our analyses.

**Measuring Openness and Unevenness**

We develop two measures: an aggregate index of openness that captures the extent that a country's economic sectors are embedded in the international economy, and a measure of upstream domestic embeddedness at the sectorial as well as aggregate level of a country. The notation we use in our formulas below refer to the Input-Output schematic provided in Table 1 above.

*Openness*

The first aspect of a national economy that we want to capture is to what extent its sectors are embedded/interconnected with the outside (non-domestic) economy. From the perspective of an individual economic sector within a country, we would like to know to what extent this sector engages in exchange with other sectors and final uses within the national borders vis-à-vis sectors and final uses outside the country, as such capturing the "neutrality" of national borders.

Among the different metrics that exist for capturing this aspect and their categorization into measures of, respectively, trade volumes and trade restrictions (see Yanikkaya 2003), we are thus interested in the former type of measuring the openness of an economy. One, if not the most, common metric is the aptly called openness index, which simply reflects the share of total imports and exports divided by gross domestic product. However, as our focus is on the level of economic sectors and the flows to and from these, we replace the GDP denominator in the more traditional openness index with the sum of domestic sectorial flows, whether for intermediate input to other sectors or for final domestic use and consumption. This also implies that our metrics are self-contained, only using data as obtained from the national Input-Output tables. As our interest lies in the connectivity *between* different sectors and as intra-sectorial flows reasonably could depend on the fragmentation and size of industrial units and companies within a sector, we consistently exclude intra-sectorial flows from this, as well as the other, metrics in our study[4].

With reference to the various parts of the national Input-Output tables (see Table 1), we apply a measure of economic integration – openness – as follows:

$$openness = \left(\sum_i^N (e_i + i_i) + \sum IFU\right) / \left(\sum_i^N \sum_{j, j \neq i}^N z_{i,j} + \sum DFU\right)$$

---

[4] The magnitude of intra-sectorial flows might be influenced strongly by concentration of firm sizes. If a sector is represented by few, or only one large firm, intra-firm flows might not get reported to statistical agencies.

The openness for a particular country and year is thus calculated by summing up all exports and imports, whether for intermediate or final use, and subsequently dividing this with the sum of all domestic intermediate and final use flows.

*Sectorial Upstream domestic embeddedness (UDE)*

Whereas the above metric captures the extent to which a national economy and its economic sectors are embedded in international production structures, the usefulness of the openness index proposed above is to capture a state of economic integration at an aggregate level. Supplementing this, we propose a metric that captures the interplay between the exports of a sector and its degree of its sourcing of domestic intermediate inputs. We operationalize this index of *upstream domestic embeddedness* (UDE) for a sector by first calculating the share of total inter-sectorial[5] domestic inputs that feeds this particular sector, subsequently subtracting the share of total exports for this sector, thus yielding the percentage-point difference between shares of domestic inter-sectorial inputs vis-à-vis foreign exports.[6]

$$UDE_i = \frac{\sum_j^n z_{j,i}}{\sum_j^n \sum_{k,k \neq j}^n z_{j,k}} - \frac{e_i}{\sum E}$$

The UDE of a sector is thus calculated as the difference between two terms: the first is the share from domestic upstream flows (the sum of all domestic inflows to a sector from all other domestic sectors divided by the total sum of all inter-sectorial domestic intermediate flows). The second term is the share from all exports (export from this sector divided by the sum of all exports).

A sector with a negative UDE value thus means that its share of total exports exceeds its share as a receiver of domestically produced inputs, whereas a positive UDE value indicates the inverse relationship. We can thus expect that the economic sectors that are inherently oriented to the domestic intermediate and final consumption markets (e.g. construction, utilities, retail sectors, education etc), have positive UDE metrics. Whereas a negative UDE value indeed indicates a sector whose significance as an exporter exceeds its share of total domestic inputs, a better understanding of the particularities of such a sector has to take foreign sectorial imports into account as well. If its share of imports are relatively low, its high share of total export values (i.e. negative UDE values) would reflect a value-producing sector that is in need of relatively few intermediate inputs, whether domestic or foreign. A sector with low (negative) upstream domestic embeddedness with a relatively high share of foreign inputs would however characterize a sector that merely acts as a link to transnational production structures, where such a position in global value chains evidently is not dependent on, or results in fewer, domestic inter-sectorial upstream linkages. To capture this distinction, the country-sector profiles we provide in our case studies below combine the UDE metrics with sectorial shares of, respectively, total exports and imports.

---

[5] Similar to the openness index and based on the same reasons, we have chosen to exclude the intra-sectorial flows in the diagonal of the Input-Output tables.

[6] A corresponding index for downstream domestic embeddedness is conceivable, i.e. where a sector's share of domestic inter-sectorial output is contrasted with its share of imports. In agreement with the contemporary literature on international political economy, testing such a corresponding downstream index in our analysis, we did however find that the most interesting findings stemmed from looking at exports vis-à-vis domestic inputs, i.e. reflecting where most of the contemporary literature on international political economy and world-system analysis puts its focus.

*Unevenness*

Although a near-zero UDE value reflects a balance between a sector's share of exports and share of total domestic inter-sectorial inputs, it is to be expected that the UDE values for the more domestically oriented economic sectors are positive across all the board. Similarly, we can expect certain sector-specific biases in the negative UDE values that we find for more export-oriented sectors.[7]

Acknowledging such sector-specific characteristics and incorporating the sectorial variance we can expect from this, the aggregate (country-wide) measure of integrational unevenness that we propose captures this variance as the sum-of-squares of the sectorial UDE values – see formula below.

$$unevenness = \sum_{i}^{N} UDE_i^2$$

Figure 2 below depicts an example economy consisting of 4 domestic sectors with domestic inter-sectorial flows as well as imports and exports to the different sectors. Excluded from this figure are intra-sectorial flows as well as domestic value-added and flows for domestic final use. The openness measure for this example is exactly equal to one (there are 200 domestic flows and 200 outside flows). With an aggregate (country-wide) UDE value of .236, the corresponding sectorial UDE values are found next to the example IO table below.

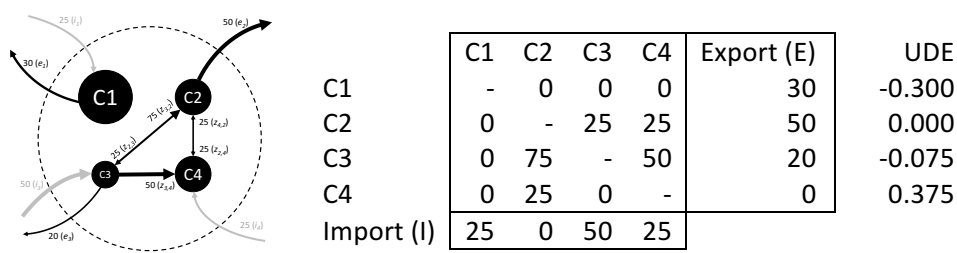

|            | C1 | C2 | C3 | C4 | Export (E) | UDE    |
|------------|----|----|----|----|------------|--------|
| C1         | -  | 0  | 0  | 0  | 30         | -0.300 |
| C2         | 0  | -  | 25 | 25 | 50         | 0.000  |
| C3         | 0  | 75 | -  | 50 | 20         | -0.075 |
| C4         | 0  | 25 | 0  | -  | 0          | 0.375  |
| Import (I) | 25 | 0  | 50 | 25 |            |        |

Figure 2: Example for calculating upstream domestic embedding

In this example, the upstream domestic embeddedness of sector C2 can be interpreted as "balanced": with half of total inter-sectorial inflows going to C2, this is matched by half of all exports originating from C2. Sector C4 reflects a domestically oriented sector: lacking exports in this example data, the domestic intermediate inflows result in a positive UDE index. Finally, sectors C1 and C3 have negative UDE values, implying that their significance as exporters exceed their significance as destinations for domestic inputs. Whereas C1 in our example lacks domestic upstream ties altogether, C3 does have upstream domestic linkages. Whereas the domestic inputs to C3 represents 1/8 of all domestic inter-sectorial inflows, its share of total exports is slightly higher (1/5).

---

[7] Although a benchmark approach could be used here, i.e. determining an average sectorial domestic/foreign ratio using all countries and years and subsequently adjusting the UDE metric to this benchmark, we preferred allowing for these inherent sectorial properties to shine through in our results, especially as our interest lies in longitudinal change.

Whereas C1 in this example constitutes an enclave sector – a cathedral in the desert – the above example demonstrate the necessity of also looking at sectorial imports for drawing such conclusions. As ¼ of all sectorial imports goes to sector C1, this indeed indicates a sector obtaining intermediate inputs from foreign, rather than domestic, sources. Similarly, foreign intermediate inputs to C3 are twice that of domestic inputs, which also has to be taken into account when interpreting the state of the sector. However, if the sectorial imports to C1 were to be zero in our example, the interpretation of its role in international production structures would be somewhat different: it could then possibly indicate a resource node at the top of the global streams of production or simply a self-sustained "cornucopian" sector that produces and exports value without needing any significant inputs, domestic nor foreign.

*Dependency*

Whereas our previous metrics examine the intermediate (inter-sectoral) flows within respective national economy, our measure of dependency measures sectorial export partner concentration. A national economy experiences higher constraint, if the export from its sectors is concentrated. We measure concentration by the relative size of the first and second largest export partner sector, where size is measured as the proportion of exports leaving the source sector. If, for example a sector exports to ten partners, 10% of all exports to each, then our dependency variable equals zero. This variable also equals zero, if a given sector exports to two partners, 50% of all exports to each. In both of these situations our source sector can avoid being dependent on a dominant target sector, by having equal size alternative partner sectors to ship to. If however, a sector exports 50% of its output to one target sector, and the second target sector takes up only 10% of exports (and the remaining 40% of exports goes to partners with even smaller shares), then this sector is dependent on a major target partner. We calculate dependency of a national economy as the mean dependencies of sectors. Sectoral dependency is measured as the difference between the largest and second largest normalized export element:

$$dependency = \frac{\sum_{i=1}^{n}(e\prime_{ij(n)} - e\prime_{ij(n-1)})}{n}, \text{ where } e\prime_{ij} = \frac{e_{ij}}{\sum_{j=1}^{n} e_{ij}}$$

Where *j* denotes all foreign sectors, in every foreign economy.

**Openness and Unevenness at the Regional Level**

The openness of European economies has increased considerably from the mid-nineties to the end of the first decade of the two thousands. Figure 3 shows the trends of our sectoral openness score by three regions. All three regions – the European core, the GIPS countries on the Western and Southern periphery, and countries in the East – followed the same basic trajectory: From an openness score of about 1 in 1995 (where the size of exports plus imports is the same as the amount of domestic inter-sectors flows) all three regions reached an openness score of about 1.5 by 2011 (with outside flows 50% larger than domestic flows). The two peripheral regions (GIPS and East) were slightly more open throughout, and the 2009 crisis shows up as a drop in the openness of all regions, but the overall trend is increasing openness.

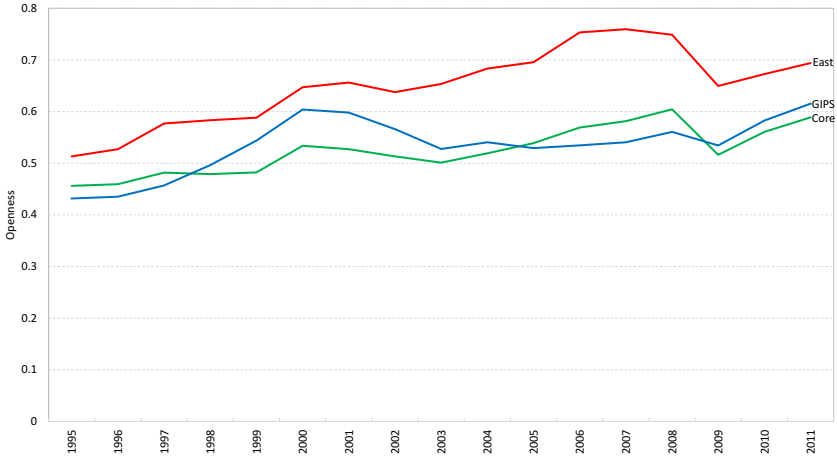

Figure 3: Mean openness of sectoral flows for three European regions between 1995 and 2011.

While there is no difference in openness across regions, is there a difference in unevenness? Our hypothesis is that it is not the extent of openness that distinguishes the European Core from its peripheries, but the domestic embedding of export sectors. In other words, it is the Core that has the capacity to benefit from this increased openness through the indirect increase in demand for the outputs of sectors feeding the export sectors.

How much does an increase in openness go together with an increase in unevenness? To estimate the overall relationship, we pool our country-year observations, where each country-year data point has a value for openness, and also a value for unevenness. This pooled time series dataset consists of 408 observations: one for each of the 17 years for the 24 countries. We use this pooled data to estimate a simple regression slope for each of the three regions. The results are visualized below on Figure 4.

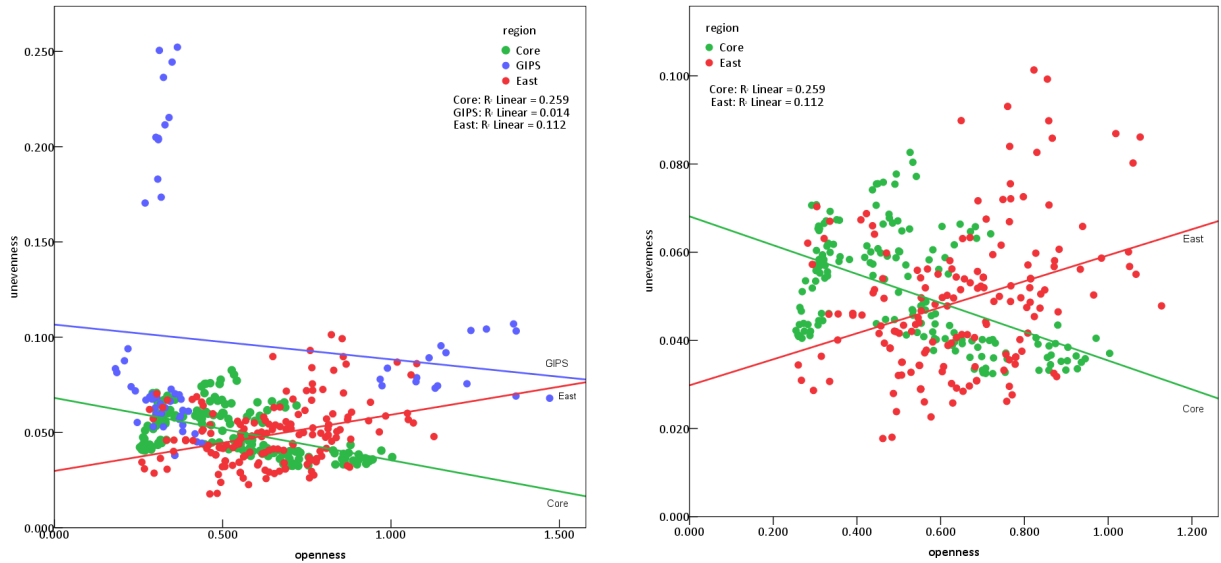

Panel 1: all three regions                                         Panel 2: Core and East regions

Figure 4: Openness and unevenness by regions, with bivariate linear regression predicted values.

The first panel of Figure 4 shows all three regions. A group of outlier data points with high unevenness and relatively low openness stands out: these are data points for Greece – the most extreme case of unevenness in terms of the lack of upstream domestic embedding of export sectors. Another group of outlier data points are to the right: with less extreme values of unevenness, but very high values of openness. These points represent Ireland: one of the most open national economies in the world. (We will discuss the trajectory of Greece and Ireland in detail later.) The regression slopes vary: for the Core and GIPS region the slope is negative: more openness results in less unevenness. The predicted line for the East has a positive slope: more openness means more unevenness here. The second panel only shows the Core and East data points. At higher levels of openness (above .80) the difference between the unevenness of Core and East economies becomes dramatic. However, there is considerable scattering in the data, and the variance of unevenness explained by openness is 26% for the best fitting region (the Core).

To test the statistical significance of the difference in the way openness is related to unevenness across the three regions we study, we employ a pooled time series regression model. The dependent variable is unevenness, and the independent variables represent regions, and the varying slopes of openness within regions; plus controls. The first variable that we include is *Openness*, to control for an overall slope between openness and unevenness. We control for a simple trend in unevenness by including a *Year* variable, which is equal to one for 1995, and goes to 17. We include an interaction between year and openness to test for a changing overall relationship between openness and unevenness. We add the total size of sectoral flows to represent the size of the economy (in thousand billion US dollars). Larger economies might be less uneven, and for small economies unevenness might be a greater risk. We include binary indicators for the three regions – the model includes GIPS and East regions as predictors, and the Core region as the omitted category. We then include interactions between the region indicators and openness – these are the variables that we are really interested in. We include a GIPS * Openness and an East * Openness interaction, and the Core is again the omitted category. We ran an ordinary least squares model, but computed one-sided p-values using a permutation test[8].

|  | Unevenness * 100 | | |
|---|---|---|---|
| Independents: | B | beta | p-value |
| Intercept | 7.656 |  | .052 |
| Openness | -4.420 | -.327 | .010 |
| Year | 0.168 | .253 | .001 |
| Year * Openness | -0.161 | -.201 | .119 |
| Size | -0.486 | -.219 | .001 |
| Region |  |  |  |
| GIPS | 2.054 | .235 | .002 |
| East | -5.709 | -.863 | .000 |
| Region interactions |  |  |  |
| GIPS * Openness | 3.741 | .291 | .019 |
| East * Openness | 8.280 | .867 | .000 |
| N | 408 |  |  |
| Adj. R-square | .336 |  |  |
| F | 26.815 |  |  |
| p-value | .000 |  |  |
| Replications | 10 000 |  |  |

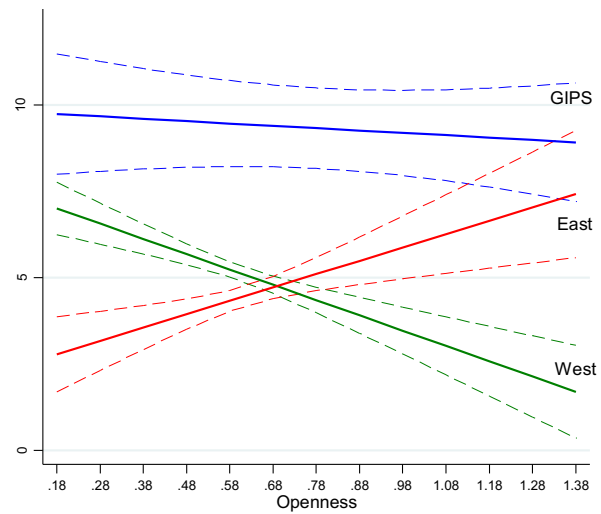

a: regression coefficients          b: marginal effects plot

Table 2: Pooled time series regression model predicting unevenness.

The model shown on panel a of table 2 indicates that overall openness is negatively related with unevenness. There is a significant time trend: unevenness increases for the entire set of European economies with time. We did not find evidence for a changing relationship between openness and unevenness with time (the interaction between time and openness is not significant). Turning to regions, the initial level of unevenness in the East is below both the level of unevenness in the Core and in the GIPS countries. (This finding is confirmed by a simple analysis of variance for unevenness across regions; with p<.050.) Our pooled time series model indicates that the relationship between openness in the East is significantly different from the same relationship in the Core region. In the East region more openness means more unevenness, at the p=.000 level of significance. We visualize the predicted levels of unevenness by region on panel b of Table 2. This marginal effects plot shows the predicted values of unevenness at various levels of openness by region, while all other variables are kept constant at their mean values. The positive relationship between openness and unevenness is specific to the East region only. The differences between the regions in the way openness relates to unevenness is not due to the overall trend of increasing openness, or the size of economies, or simple random noise. National economies in the East are becoming more uneven with increasing openness, compared to Core economies.

---

[8] Permutations tests for the p-values of coefficients is especially appropriate since the observations are not drawn as a sample from a large population, but represent all cases – all the country years in the period we consider (Good 2006).

**Openness and Dependency**

After showing evidence for the regional differences in how openness is related to unevenness, we analyze similar relationships regarding dependency. As economies in the East are becoming more uneven, are they also becoming more dependent as well? To answer this question, we constructed the same regression model that we used for unevenness. The dependent variable here is dependency – the mean difference between the largest and second largest export partner for domestic sectors. The independent variables are the same: *Openness*, *Year*, an interaction term between *Openness* and *Year*, the overall *Size* of the economy, and indicators of the three regions, plus interactions between *Region* and *Openness*. Here, again, standard errors are estimated by a permutation test.

|  | Dependency ∗ 100 | | |
|---|---|---|---|
| Independents: | B | beta | p-value |
| Intercept | 1.532 |  | .110 |
| Openness | 4.441 | .301 | .001 |
| Year | 0.493 | .677 | .000 |
| Year * Openness | -0.511 | -.582 | .000 |
| Size | -0.699 | -.288 | .000 |
| Region |  |  |  |
| GIPS | -.227 | -.024 | .425 |
| East | -1.808 | -.250 | .107 |
| Region interactions |  |  |  |
| GIPS * Openness | 2.138 | .152 | .140 |
| East * Openness | 4.923 | .471 | .013 |
| N | 408 |  |  |
| Adj. R-square | .281 |  |  |
| F | 29.980 |  |  |
| p-value | .000 |  |  |
| Replications | 10 000 |  |  |

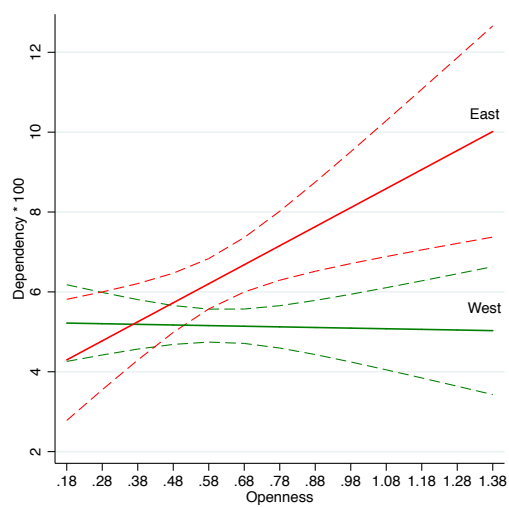

a: regression coefficients          b: marginal effects plot

Table 3: Pooled time series regression model predicting dependency.

As the first panel of Table 3 shows, dependency of an economy is positively related to its Openness; dependency increases with time, but the impact of Openness on dependency is mitigated with time. Larger economies are less dependent on average. Regional difference is only manifest in the added impact of Openness on Dependency in the East. The second panel of Table 3 shows the differences in how Openness is related to Dependency, by regions. For clarity, the GIPS region is omitted (it falls between the fitted lines of East and West). The marginal effects plot – keeping all independent variables but Openness constant at their means – shows that there is no significant relationship between openness and dependency in the West, while in the East increase in openness goes together with increase in dependency. The GIPS region shows a pattern that is in between the Core and East. The trend in the GIPS region is similar to the East – more openness goes together with more dependency. However, this trend is not statistically significant (p=.140), but the sign is positive.

**Openness and Unevenness by Countries**

After analyzing openness and unevenness at the level of regions, now we analyze this relationship at the level of national economies. Openness increases in all three regions, and openness related differently to unevenness within regions. Now we turn to the variation at the level of national economies. As Figure 5 shows, the general tendency is increasing openness for all but two of the 24 economies. Countries in the Core region are relatively bounded in their increase of openness, with increase in the 5% - 25% range. The East and the GIPS regions are more diverse. In the GIPS region the openness of Ireland increased dramatically, by 50%. In Portugal, Spain and Greece the increase in openness is below European average of 0.161 (16.1% higher openness in 2011 over 1995). The highest increase in openness was in Hungary: a 67% increase. There are only two countries, Estonia and Latvia, where the openness in 2011 is less than the openness in 1995. Unevenness (shown on the secondary axis) varies more than openness. Some countries experienced a decrease in unevenness, while some (especially Greece) saw major increase.

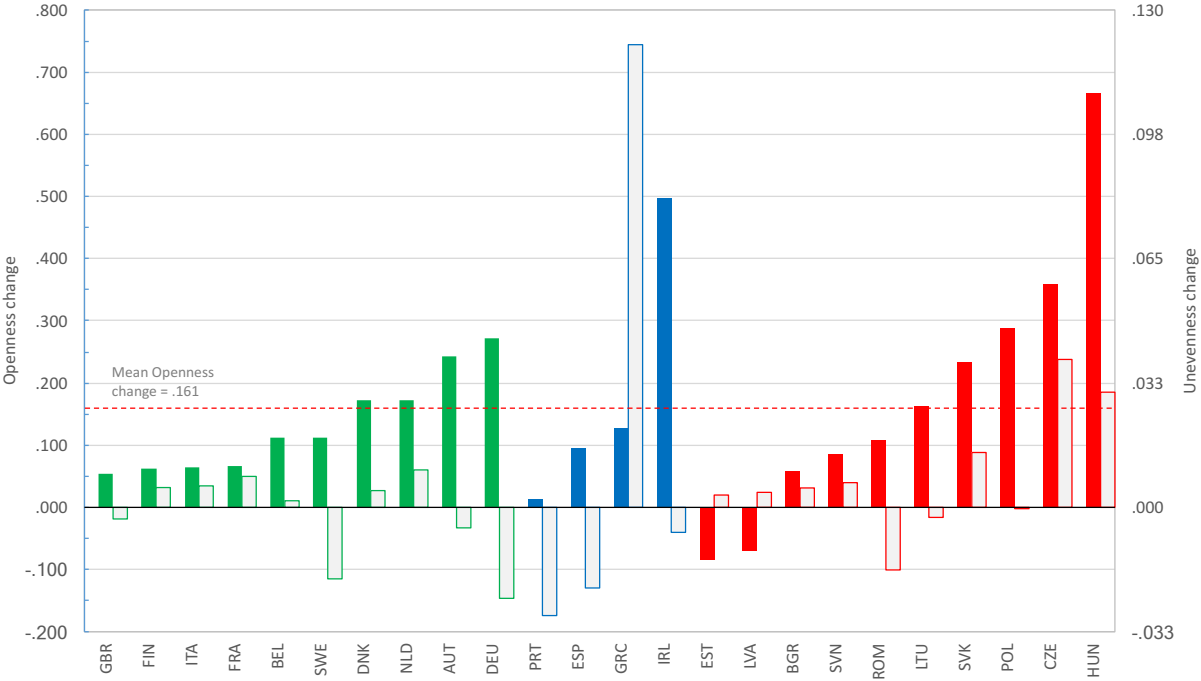

Figure 5: Relative change in openness and unevenness for national economies between 1995 and 2011.

To compare countries by the tradeoffs between openness and unevenness, we computed regression slopes for each country. While in the previous section we estimated the differences in the openness-unevenness association by region, here we do so at the country level. As the datasets become small (17 observation in each case), we only use bivariate models. We test for the significance of the slope coefficient using a permutation test again, especially appropriate for small samples. The results of the regression estimates are presented in Figure 6.

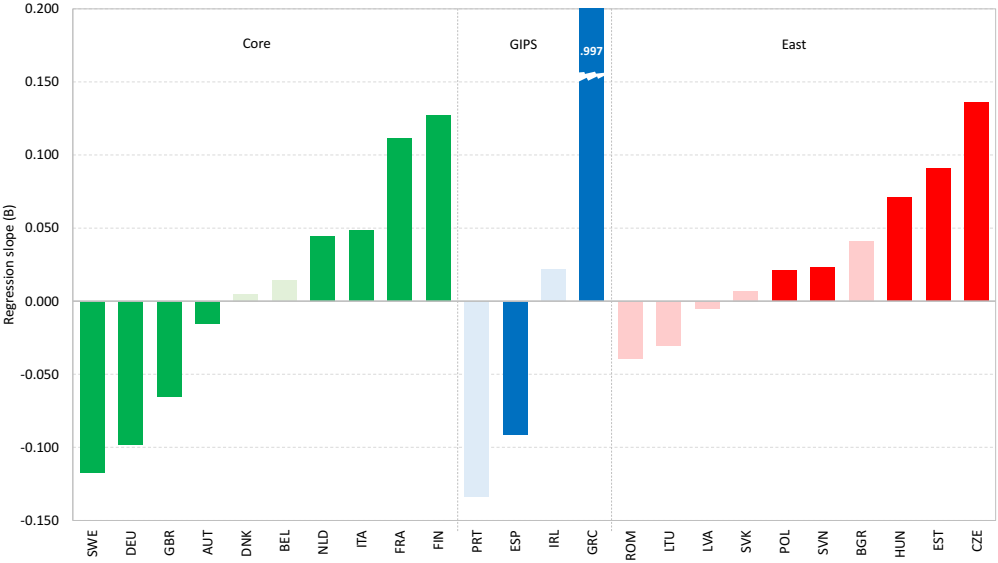

Figure 6: Regression slopes (B coefficients) for predicting unevenness based on openness. Light shading indicate statistically insignificant slopes (p>.010).

Negative slope coefficients mean that when the openness of an economy increases, the unevenness decreases. In such countries the increase of exports, for example, is located in sectors that are already well embedded in domestic upstream flows. A positive slope means that when the openness of that economy increases, unevenness also increases. In such an economy, for example, sectors that increase their exports are disembedded from the domestic inter-sectoral flows.

Figure 6 shows that the three economic regions are not homogenous, nevertheless, the overall differences among regions seen before are manifest in the country breakdown as well. The most important inequality is between the Core countries and the countries in the East: Four of the core countries (Sweden, Germany, Great Britain, and Austria) show a significant negative relationship between openness and unevenness, while none of the East countries have a significant negative coefficient. (Only one none-core country has a negative coefficient: Spain.) What this indicates is that several core countries have increased their openness in a way that benefits domestic upstream embedding of their sectors, while there is evidence for the opposite trend in the East. We find five economies in the East (Poland, Slovenia, Hungary, Estonia, and Czech Republic) where an increase in openness means dis-embedding from domestic upstream flows.

Structural benefits of increasing openness seem to accumulate in the Core. But these benefits are not experienced by all core economies: there are four of them (Netherlands, Italy, France, and Finland) where openness brings unevenness (domestic dis-embedding).

**Trajectories**

Up to this point we considered only incremental (annual) change, or overall change from 1995 to 2011. In this section we consider the shape of trajectories that economies traveled in the space of openness and unevenness.  As our initial motivation was to distinguish between transient and durable inequalities, we need to know the historical shape of changes.  We construct trajectories charts for each country aggregating our data into three year periods for smoothing. We argue that the concept of trajectory is especially relevant to understand economic development in the space of openness and unevenness. While in the previous analyses we were identifying overall linear trends, here we are interested in non-linear developmental paths that are specific to individual countries, or types of countries.  Our trajectory charts show the traces for each country colored by the slope of the trajectory.  A red line indicates that the trajectory follows a statistically significant positive linear trend (unevenness increases with openness).  A blue trajectory has a significant negative trend, while a grey trajectory has no significant linear trend.

*The Core*

Figure 7 presents the trajectories of countries in the space of openness and unevenness from 1995 to 2011.  Great Britain, Germany, Sweden, and Austria have significant negative slopes: their unevenness decreases with increasing openness.  Of the trajectories in the Core, it is Germany that shows the longest distance traversed – the greatest increase in openness, and the greatest corresponding decrease in unevenness.

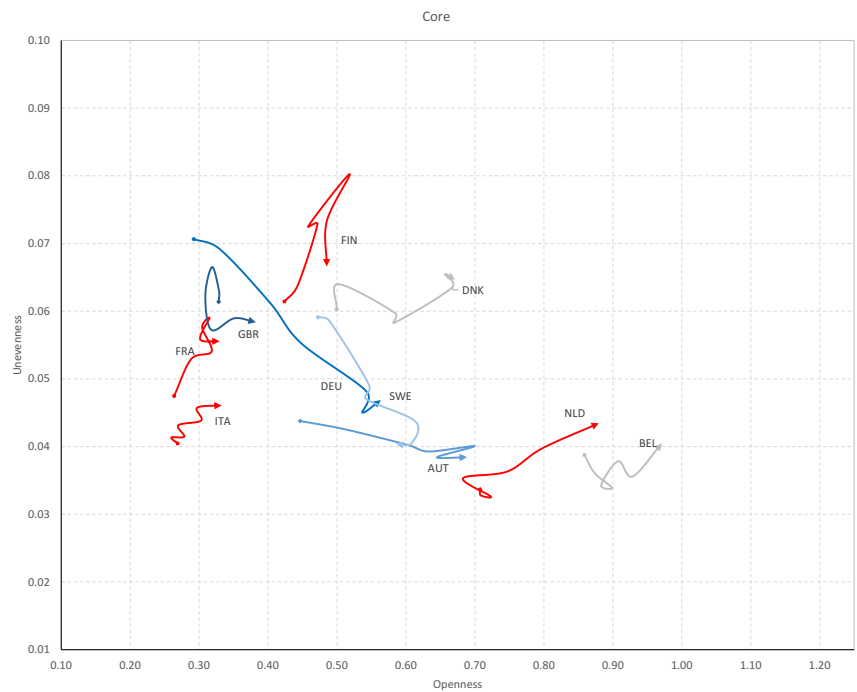

*Note*: Color indicates slope: Blue hues: significant negative slope; Gray: insignificant slope; Red hues: significant positive slope.

Figure 7: Core country trajectories in the space of openness and unevenness.

*Germany*

What happened in the German economy that explains the increase in openness that at the same time decreased unevenness? To understand this, we analyze the starting and ending point of this trajectory at the level of particular German sectors. Figure 8 shows German sectors in 1995 and in 2011, ranked by their upstream domestic embedding. For each sector we also show the export share and the import share (the proportion of the economy's imports and exports that happened in that sector). Sectors at the left have the lowest values of domestic upstream embeddedness, and sectors on the right have the highest values. An economy where the line representing upstream domestic embedding is completely flat is an economy that is perfectly balanced: for every sector the share in exports is the same as the share in domestic intermediate inputs. This line for Germany in 1995 is not flat: on the left hand side the transport equipment sector has the highest negative value: -0.128. This means that the share of this sector in the total domestic inter-sectoral flows (0.058, or 5.8%) is 0.128 less than the proportion of all German exports that is located in this sector (0.186, or 18.6%). On the right end of the chart there is the construction sector, with a positive upstream embeddedness score. The construction industry consumed 11% (0.110) of domestic inter-sectoral flows, while its export share was only 0.3% (0.003). Thus its' UDE score is 0.107. For the following case studies we present only the bottom five sectors in the ranking by upstream domestic embeddedness – the most dis-embedded export sectors.

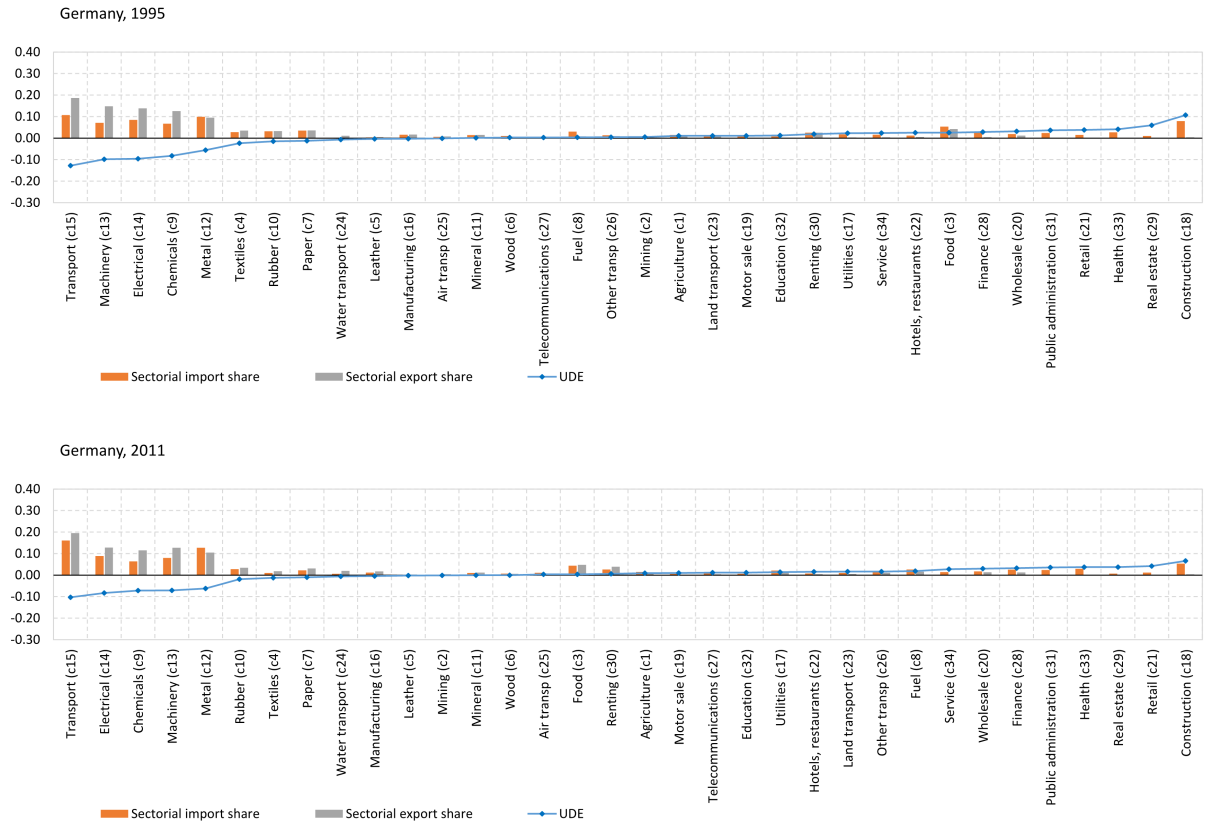

Figure 8: German sectors in 1995, ranked by upstream domestic embedding (UDE).

As we have seen on the chart of Core country trajectories, Germany's openness steadily increased from 1995 to 2011, while its unevenness has decreased. Comparing the sectoral breakdowns from 1995 and 2011 there is no apparent radical difference at first sight. The charts tell of important changes though: key sectors increased both their openness (mostly import shares), and both their domestic intermediate inputs – their upstream embedding.

The key example is the leading sector of Germany, the transport equipment manufacturing sector. This sector has increased its export share between 1995 and 2011 from 18.6% to 19.5%, while it also increased its share from domestic sectoral inputs from 5.8% to 9.2%. Even though this period saw a great increase in the foreign production and value added component in this sector, there was a great increase in German inputs as well. The inputs of the German metallurgy sector (the largest domestic supplier) to the transport equipment sector increased by 143.1%, while the overall increase of the German economy (measured in total flows) was 47.9%. Supplies from the renting sector has increased by 244.0% (reflecting the major increase in the practice of relying on rented equipment in industry). Flows from German wholesale increased by 107.4%.

Similar trends can be observed in the other major German sectors as well: machinery, electronics, chemical, and metallurgy sectors. The machinery sector increased its inputs from domestic metallurgy by 97.5%, from renting by 123.4%. The metallurgy sector increased inputs from renting by 102.2%, from utilities by 74.7%. In sum, the German economy managed to both increase its production abroad, and to increase its reliance on domestic sectors. Why and how this was possible is a question outside the scope of this article – but one might guess that technological change (the increasing significance of powdered metals), labor policy (pacts to curb domestic wage increase), and dependence on high quality specific inputs might constitute parts of the causes.

*France*

The trajectory of France is orthogonal to the German trajectory: the French economy has been becoming more open, but it has also became more uneven in the process. The charts on Figure 9 show the five sectors with the least domestic upstream embeddedness in 1995 and in 2011. (For the sake of saving space we use only the most disembedded five sectors for this and following country cases.) The chart from 1995 looks very similar to the same chart from Germany – with the same sectors, and with slightly smaller negative values for embeddedness. By 2011 the list and order of these top five sectors remain the same, with an increased dis-embedding.

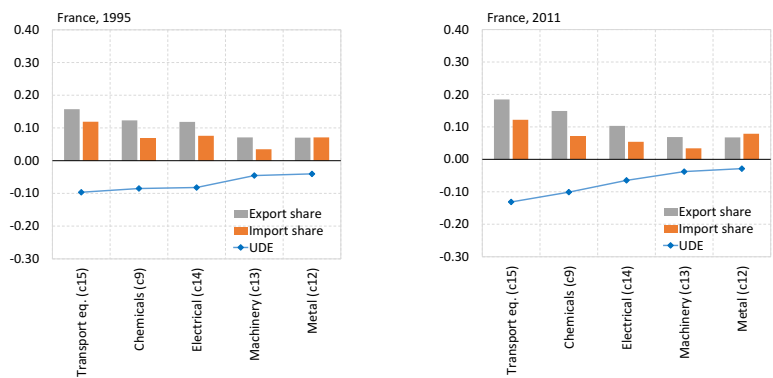

Figure 9: French sectors in 1995 and 2001, bottom five sectors, ranked by upstream domestic embedding (UDE).

Similar to Germany, the sector with the largest share of exports is the transport equipment sector. In 1995 this sector was responsible for 15.7% of all exports, while it used 6.1% of domestic intermediate flows. In 2011 the export share of the export share of this sector increased to 18.5%, while the share in domestic upstream decreased to 5.3%. The composition of domestic inputs to the transport equipment sector changed little – with equipment renting, metallurgy wholesale, and electrical equipment sectors being the top suppliers.

The second largest sector, chemicals shows a similar pattern: its export share increased from 12.3% to 14.9%, while the share from domestic inputs increase at a slower rate (from 3.8% to 4.8%). There was practically no change in the outside and inside flows of the other sectors in the top list (electrical, machinery, and metallurgy). Overall, it seems that France was not able to involve domestic supply sectors into the process of increasing transnationalization in the same way that Germany was able to accomplish.

*Trajectories in the GIPS countries*

Figure 10 shows the country trajectories in the space of openness and unevenness for the GIPS countries. Greece is a clear outlier in its extreme increase in unevenness, with only a moderate increase in openness. Ireland is the opposite of the Greek story – outstanding increase in openness with only moderate levels of unevenness.

*Greece*

The Greek economy in 1995 was not significantly more uneven than the German economy in the same year (Greek unevenness was 0.083, German unevenness was 0.071). By the end of the time period we study the Greek economy became the most uneven (unevenness = 0.204), while Germany's unevenness decreased to 0.047. What happened?

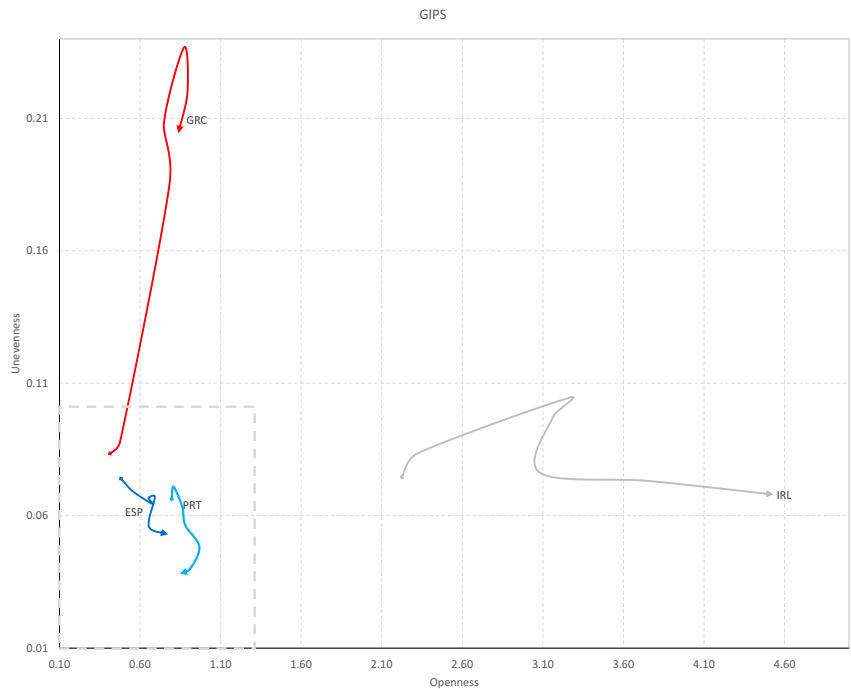

*Note*: Color indicates slope: Blue hues: significant negative slope; Gray: insignificant slope; Red hues: significant positive slope. The area outlined by dashed grey rectangle represents the area covered by the Core and East trajectories charts.

Figure 10: GIPS country trajectories in the space of openness and unevenness.

The sectoral breakdown of the Greek economy is presented on Figure 11. In 1995 the moderate level of unevenness in upstream domestic embedding was chiefly due to water transport, textiles, agriculture and metallurgy. In this year water transport was responsible for 16.0% of exports, textiles and agriculture represented about 12% of exports each, and metallurgy was 8.6%. By 2011 the openness of the Greek economy increased only slightly (see Böwer et al 2014 for a report on the missing Greek exports), but the sectoral structure of exports changed drastically. Water transport dominated Greek exports, with 42.1% of all exports originating from this sector. The main contributor to the greatly increased unevenness is the water transport sector. Greece is the most important player in maritime transport in the world, it controls 16.2% of global water transport capacities. As the export share of water transport increased dramatically (from 16.0% to 42.1%), its share in domestic inputs did not follow this increase (share in domestic intermediate inputs increased from 2.1% to 9.1%). In absolute nominal terms the amount of exports from the water transport sector increased from 1527 million USD to 17 905 million USD (an eleven-fold increase), while domestic intermediate inputs from other sectors increased from 631 million USD to 1952 million USD (a threefold increase). The main domestic inputs to this sector come from other transport services, which include cargo handling, storage, and transport agency services.

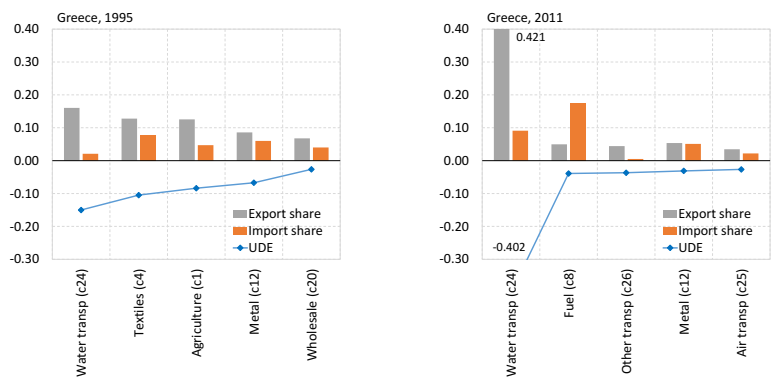

Figure 11: Greek sectors in 1995 and 2001, ranked by upstream domestic embedding (UDE).

*Ireland*

In many respects Ireland is the exact opposite of Greece: The trajectory of Ireland is dominated by increasing openness, with a first phase where unevenness increases slightly (between 1995 and 2001 unevenness increasesd from 0.075 to 0.107), and a second phase where unevenness stays around 0.070 (from 2002 to 2011).

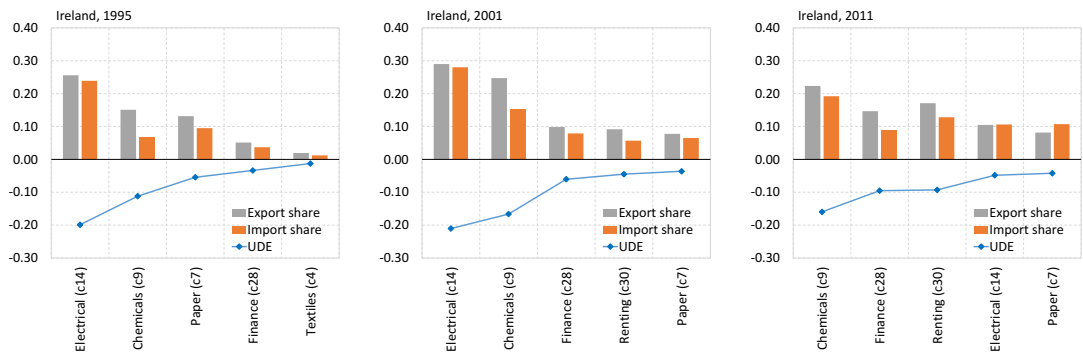

Figure 12: Ireland sectors in 1995, 2001 and 2001, ranked by upstream domestic embedding (UDE).

The major growth of the Irish economy from the nineties was fueled by a great influx of foreign direct investmemnt into high technology manufacturing and services (Kirby 2010, Kirby at al 2010). Over the years of the nineties and two thousands Ireland became a hub for electronics companies. As Figure 12 shows, electronics was the largest contributor to the unevenness of the economy: it contributed to 25.6% of exports, while it used only 5.7% of domestic intermediate inputs, and it used 19.8% of all imports. The unevenness of the economy increased slighlty to 2001, with the electronics sector further increasing its export share to 29.0%, with a 7.9% share of domestic intermedaite inputs, and the chemical industry drastically increasing its export share from 15.1% to 24.7%, while its domestic usptream share increased from 3.9% to 8.1%.

The unevenness of the Irish ecnomy decreased from 2001 to 2011, partly due to changes in the weight of sectors, and partly due to changes within sectors. The export share and unevenness of the electronics sector decreased drastically: the export share decreased from 29.0% to 10.5%, while the setor's share from domestic inputs declreased much less (from 7.9 to 5.6%). Overall, the distribution of export shares become more even, and chemical products became the top export share sector. The nature of the electronics sector seems to have changed as well: whreas in 1995 this sector mostly used inputs from wholesale and retail, by 2011 the weight of equipment renting, chemical products, metallurgy, and utilities increased significantly. The sector switched from simple assembly to a deeper itnegration with domestic sectors, and re-oriented towards healthcare equiment manufacturing. In sum, the Irish trajectory switched from a parallel increase of openness and uneveness to a trajectory where drastic further increase in openness was paired with a marked decrease in unevenness.

*Trajectories in the East*

Trajectories in the East region of the EU are more varied and complex, than the trajectories in the Core and GIPS regions. As our pooled time series model of distance traversed indicated, East economies experienced larger jumps from one year to the next – as it is apparent on Figure 13.

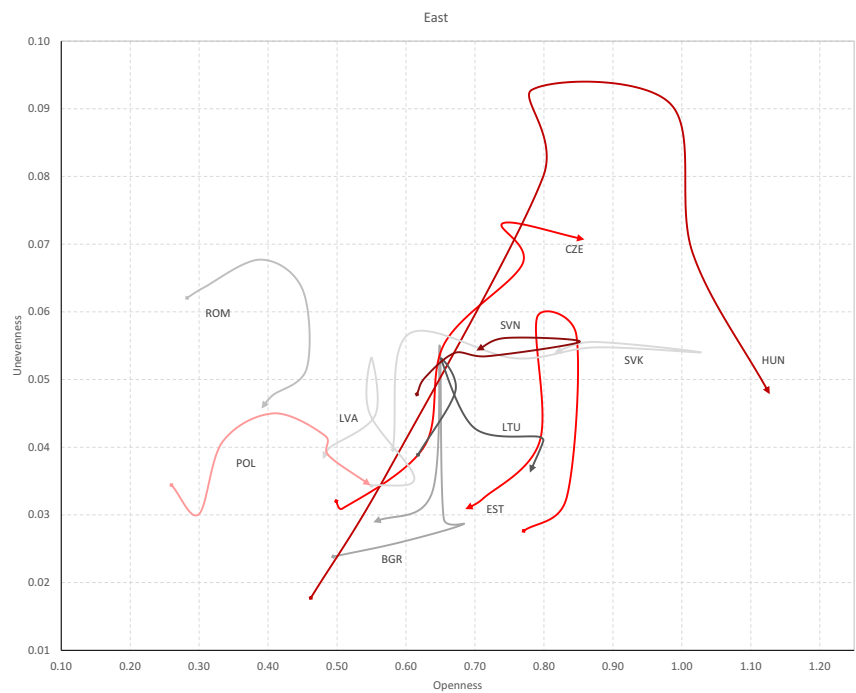

Figure 13: Trajectories of economies in the space of openness and unevenness; by region. Framed area within the GIPS chart indicates area of Core and East charts. Red hues indicate significant positive slope, blue hues indicate significant negative slopes. Gray hues indicate insignificant slopes.

The longest trajectory is the Hungarian one, with a marked increase of both openness and unevenness in the first part (between 1995 and 2004), followed by a turning point, onto a trajectory of increasing openness with decreasing unevenness. We label this a 'turning point trajectory' – a pattern followed by Poland as well (to the left of the Hungarian trajectory). There is a second kind of trajectory as well, that start with increasing openness and unevenness, and then turns back, with decreasing openness and unevenness. We label this a 'retrograde trajectory'. Figure 14 separates trajectories in the East accruing to these two patterns.

Hungary best exemplifies the turning point trajectory, with Czech Republic, Poland, and Lithuania in the same category. This trajectory suggests a structural adjustment to increasing openness, a change to a path where an increased embedding of export sectors into the domestic intermediate sectoral flows becomes possible.

Estonia is the best example of the retrograde trajectory, with Bulgaria, Latvia, Romania, Slovakia, and Slovenia also in the same category. This trajectory suggests a kind of change, where the domestic embedding improves not by increasing flows inside, but by decreasing flows with the outside world.

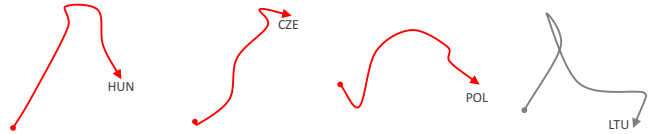
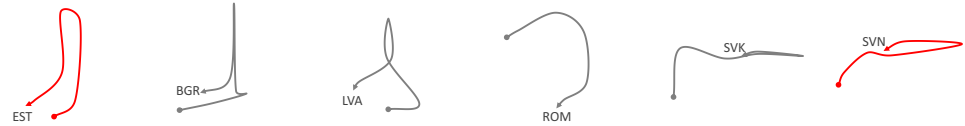

*Note*: Trajectories are not drawn to the same scale.

Figure 14: Types of trajectories of East European economies.

*Hungary*

The Hungarian economy experienced the greatest amount of change of any East European economies, in terms of openness and unevenness.  Interestingly, the most even economy in our entire dataset was the Hungarian economy in 1995.  As the first panel of Figure 15 shows, sectors in the Hungarian economy in 1995 had very similar, and evenly distributed export shares.  None of the sectors had an export share exceeding 10%, and the sector with the highest export share – Food – was actually over-embedded in domestic upstream flows (this sector was responsible for 9.1% of all exports, and used 14.9% of all intermediate flows).  The sector with the second highest export share – Metallurgy – was the most uneven of all sectors, with 8.5% of exports and 3.2% of domestic upstream flows.

This even sectoral structure changed drastically over the next ten years.  By 2004 the Hungarian economy became highly uneven.  The sector mostly responsible for this is Electronics: by 2004 this sector was responsible for 32.9% of all exports, used 28.4% of all imports, but only used 6.1% of domestic sectoral output.  By 2001 Hungary was responsible for half of all electronics exports from Eastern Europe.  A key example of the kind of electronics operations responsible for this is IBM Storage Products: an assembly plant of computer hard drives, started operating in 1996. By the end of the nineties this company became the second largest exporter.  In 2002 IBM decided to close the assembly operation, and consolidate hard drive assembly in Asia.  The second sector that contributed to unevenness in 2004 was the manufacturing of transport equipment.  This sector gave 13.9% of all exports, using 11.1% of all imports, and using 4.1% of domestic intermediate output.

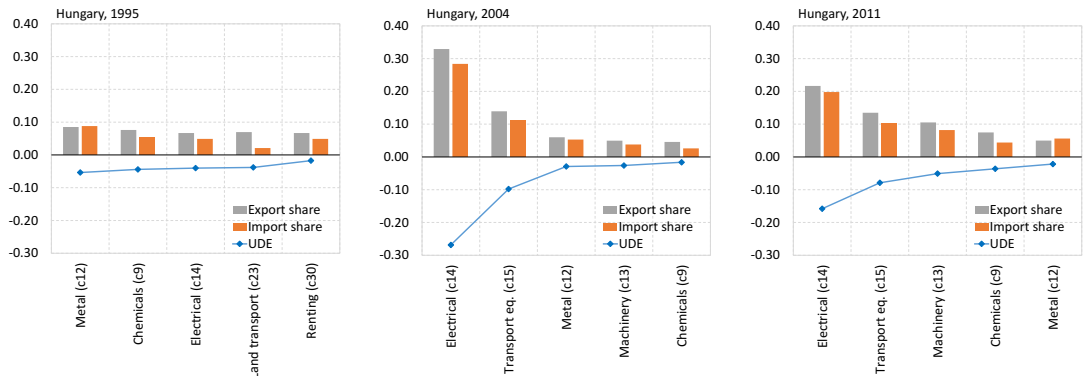

Figure 15: Hungary sectors in 1995, 2004 and 2011, ranked by upstream domestic embedding (UDE).

What was the nature of the turning point after 2004? By 2011 the list of five least domestically embedded sectors did not change much – only the ordering of the third, fourth, and fifth sectors. Electronics and transport equipment are still the sectors responsible for the highest share of exports, although their share decreased: from 32.9% to 21.7% for electronics, and from 13.9% to 13.5% for transport equipment. Electronics changed the most: the marked decrease in the export share was paired with a minor change in the proportion of domestic output used (from 6.1% to 5.9%). A sign of switching away from simple assembly is the increasing amount of output used from the domestic machinery sector. (In 2004 2.1% of all intermediate inputs to electronics came from domestic machinery sector, while in 2011 this proportion was 19.9%.) While the export share of the transport equipment sector decreased slightly (from 13.9% to 13.5%), its share in domestic inputs used increased (from 4.1% to 5.6%). Much of this increase was due to the rise in the share of the machinery sector out of all domestic sectoral inputs (this share increased from 9.5% to 40.0%).

*Estonia*

Estonia is an example for a retrograde trajectory – an economy that started to become more open and uneven, and then both openness and unevenness declined. Estonia was seen as a key example of an institutional restructuring success in the nineties and early two thousands (Bohle and Greskovits 2012). The case of Estonia is marked by an increase in openness mostly due to FDI-led electronics production, then an extreme decline during the 2008-2009 economic crisis. In 2008 Estonian GDP declined by 5.1%, then in 2009 it declined by 13.9%.

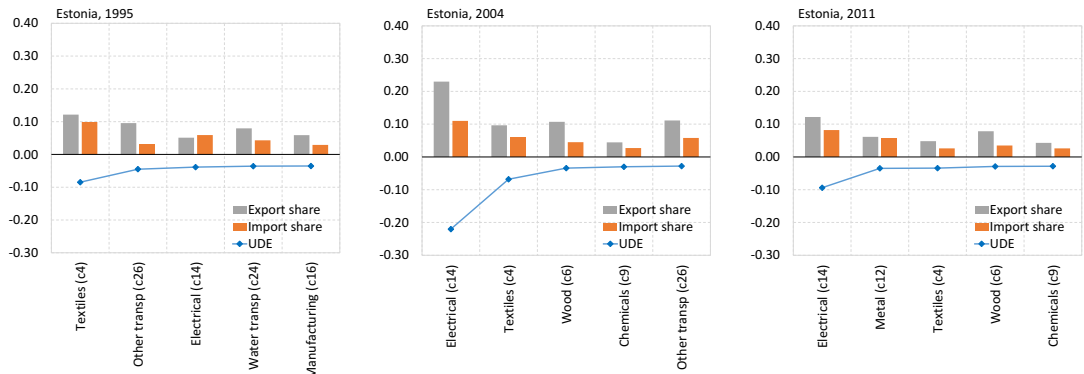

Figure 16: Estonia sectors in 1995, 2004 and 2011, ranked by upstream domestic embedding (UDE).

The increase in openness and unevenness was mostly driven by the electronics sector, as Figure 16 shows. While in 1995 this sector gave only 5.1% of exports, and used 1.2% of domestic intermediate inputs, in 2004 it gave 23.0% of exports, and used only 0.9% of domestic input. The size of the electronics sector is much below the size of Hungarian electronics – total electronics exports from Estonia were never larger than 6% of the total electronics exports from Hungary. The output of the sector came from a handful of companies (mostly from Ericsson), and depended on Scandinavian export markets. By 2011 the export share of the sector declined to 12.2%, and its share from domestic sectoral output used increased to 2.8%.

**Conclusions**

European economic integration resulted in unprecedented growth in economic openness throughout the continent. The economic crisis of 2008 and 2009 left only faint dips in the increasing trend of integration, that over a decade and a half increased by fifty percent. Yet, there has been a dramatic unevenness in the growth of economic openness, with Northern Europe, Ireland and the Visegrad group being the leaders in the internationalization of their sectoral flows and Southern Europe, Baltics and Balkans experiencing lower than average integration, with even a decrease in sectoral economic openness in Estonia and Latvia. An exclusive attention to flows between national economies would miss however a key dimension of increasing integration: the impact of openness on the domestic structure of economies and the extent to which integration is embedded in the domestic flows.

The core of the core – Germany, Great Britain, Sweden, Austria – are economies that can increase their participation in transnational flows and at the same time let their domestic sectors benefit twice: once as exporters, and twice as suppliers of other sectors that are exporters. The domestic sectoral hinterland grows with the flagship export sectors, in a virtuous circle of growth. German export sectors – primarily exporting sophisticated transport equipment, machinery, electronics and chemical products – rely on other sophisticated domestic products, with sectors innovating and learning together.

Then there are regions – Eastern Europe and Greece – where this does not seem to be the rule. In the East, as a main trend, exporting means disconnecting. Export is concentrated in sectors that are assembling imports, and rely little on other domestic sectors.  A whole sector can be built up quickly, and, because domestic inputs does not bind the sector, can be moved overnight to another continent.

Beyond focusing on trends, we also focus on trajectories.  The first finding here is that the trajectories in the East show more volatility than the trajectories in the core.  As the trajectories tell, the increase in unevenness with increasing openness in the East is not a linear trend – there are signs for turning points. Economies in Eastern Europe will not be heading down a disembedding path indefinitely – we haven't found a trajectory type for a truly vicious circle. (Maybe Greek and the Czech trajectories come close.) In the East economies are experiencing a turning point.  Partly due to opportunities opened by the 2008 crisis, export sectors are growing domestic roots, and well embedded domestic sectors start accessing export markets.  Hungary, Poland, and Lithuania seem to switch over to a path similar to the core.

Contrary to the expectations of either the optimistic vision of integration as convergence, or the pessimistic idea of structural reproduction of the peripheral status, our findings thus show that combining openness with domestic embeddedness is indeed possible, and it is not predetermined by the initial status of the economy as a core or a peripheral one. While the Hungarian or the Polish trajectories suggest that the pathway towards the core is possible in the periphery, the French trajectory suggests that core countries are not immune to disembedding either. Developmental pathways, rather than being merely the result of the structural starting conditions are rather always shaped by the developmental agency. Furthermore, we find important intra-peripheral national variation in developmental pathways, similar to the one discussed by Bohle and Greskovits (2012).  Our results thus suggest that at least with respect to trade flows, the developmental implications of increased openness may be primarily the result of domestic developmental agency, rather than the supranational one (see Bruszt and Vukov introduction to this issue on supranational developmental agency).

Further research is needed to identify the impact of unevenness on other developmental outcomes. This article aimed at highlighting an inequality that we have theoretical reasons to expect to be connected to underdevelopment, but we have not analyzed these outcomes here.  There are several hypotheses that need to be investigated – for example, is there an optimal value for unevenness that is not zero?  Are there trade-offs, where decreasing unevenness is related to, for example, increasing unemployment?  We should also investigate the limits of growing unevenness, and also decreasing unevenness.  What are the limits for the Greek economy in terms of the domestic disembedding of export sectors?  How far can the German economy develop by increasing openness and reducing unevenness?  Our goal in this article was to highlight these inequalities, using the new perspective that the WIOD dataset offers with transnational input-output flows.